\documentclass[twocolumn,showpacs,preprintnumbers,superscriptaddress,amsmath,amssymb,floatfix,aps]{revtex4}

\usepackage[dvips]{graphicx}
\usepackage{dcolumn}
\usepackage{bm}
\begin{document}

\title{The distance upon contact:
Determination from roughness profile}

\author{P. J. van Zwol}
\affiliation{Materials innovation institute M2i and Zernike
Institute for Advanced Materials, University of Groningen,
Nijenborgh 4, 9747 AG Groningen, The Netherlands}

\author{V. B. Svetovoy}
\affiliation{MESA$^+$ Institute for Nanotechnology, University of
Twente, PO 217, 7500 AE Enschede, The Netherlands}

\author{G. Palasantzas}
\affiliation{Materials innovation institute M2i and Zernike
Institute for Advanced Materials, University of Groningen,
Nijenborgh 4, 9747 AG Groningen, The Netherlands}

\date{\today}

\begin{abstract}
The point at which two random rough surfaces make contact takes
place at the contact of the highest asperities. The distance upon
contact $d_0$ in the limit of zero load has crucial importance for
determination of dispersive forces. Using gold films as an example
we demonstrate that for two parallel plates $d_0$ is a function of
the nominal size of the contact area $L$ and give a simple
expression for $d_0(L)$ via the surface roughness characteristics.
In the case of a sphere of fixed radius $R$ and a plate the scale
dependence manifests itself as an additional uncertainty $\delta
d(L)$ in the separation, where the scale $L$ is related with the
separation $d$ via the effective area of interaction $L^2\sim\pi
Rd$. This uncertainty depends on the roughness of interacting
bodies and disappears in the limit $L\rightarrow \infty$.
\end{abstract}


\maketitle

\section{Introduction}
\label{sec:1}

The absolute distance separating two bodies is a parameter of
principal importance for the determination of dispersive forces
(van the der Waals \cite{Ham37}, Casimir \cite{Cas48} or more
general Casimir-Lifshitz force \cite{Lif56}). The absolute
distance becomes difficult to determine when the separation gap
approaches nanometer dimensions. This complication originates from
the presence of surface roughness, which manifests itself on the
same scale. In fact, when the bodies are brought into gentle
contact they are still separated by some distance $d_0$, which we
call the distance upon contact due to surface roughness.

We are interested in the dispersive forces when stronger chemical
or capillary forces are eliminated. In this case $d_0$ has a
special significance for adhesion, which is mainly due to van der
Waals forces across an extensive noncontact area \cite{Del05}. The
distance $d_0$ is important for micro (nano) electro mechanical
systems (MEMS) because stiction due to adhesion is the major
failure mode in MEMS \cite{Mab97}. Furthermore, the distance upon
contact plays an important role in contact mechanics \cite{Per07},
is very significant for heat transfer \cite{Vol07}, contact
resistivity \cite{Rab95}, lubrication, and sealing \cite{Per00}.
In addition, it has also importance in the case of capillary
forces and wetting \cite{Del07,Zwo08c,Per08}, where knowledge of
$d_0$ provides further insight of how adsorbed water wets a rough
surface.

The distance upon contact $d_0$ between a sphere and a plate
\cite{Har00,Dec03} plays a key role in modern precise measurements
of the dispersion forces (see \cite{Cap07} for a review) where
$d_0$ is the main source of errors. In Casimir force measurements
$d_0$ is determined using electrostatic calibration. In this case
the force dependence on the separation is known, and one can
determine the absolute separation (see resent discussions
\cite{Kim08,Man09,Lam08}). Even when the distance is not counted
from the point of contact \cite{Kim08,Man09,Jou09} local
realization of roughness as shown in this paper will contribute to
uncertainty of the absolute separation.

Independent attempts to define $d_0$ were undertaken in experiments
measuring the adhesion energy \cite{Del05}. It was proposed
\cite{Hou97} to take $d_0$ as the sum of the root mean square (rms)
roughnesses of two surfaces  upon contact. This definition is,
however, restricted and can only be used for rough estimates as
stressed in \cite{Hou97}. Obviously, the distance upon contact has
to be defined by the highest asperities.

In this paper we propose a simple method for determination of $d_0$
from the roughness profiles of the two surfaces coming into contact.
For two plates it is explicitly demonstrated that $d_0(L)$ is scale
dependent, where $L^2$ is the area of nominal contact. We discuss
also application of our method to the sphere-plate configuration. In
this case it is shown that $d_0$ determined from the electrostatic
calibration can differ from that playing role in the dispersive
force and the difference is scale (separation) dependent.

In Sec. \ref{sec:2} we report briefly the details of our film
preparation and characterization. In Sec. \ref{sec:3} the
roughness profiles in the plate-plate configuration are discussed
and the main relation connecting $d_0$ with the size of the
nominal contact is deduced. The sphere-plate configuration is
discussed in Sec. \ref{sec:4} together with uncertainty in $d_0$.
Our conclusions are collected in Sec. \ref{sec:5}.

\section{Experimental}
\label{sec:2}

The surfaces we use in this study were gold films grown by thermal
evaporation onto oxidized silicon wafers with thicknesses in the
range $100-1600\; nm$ and having different rms roughnesses. A
polysterene sphere (radius $R=50\; \mu m$), attached on a gold
coated cantilever, was first plasma sputtered with gold for
electrical contact, and then a $100\; nm$ gold film grown on top of
the initial coating. The deposited films were of uniform thickness
and of isotropic surface morphology as was confirmed independently
with atomic force and scanning electron microscopy on different
locations.

\begin{figure}[ptb]
\begin{center}
\includegraphics[width=0.4\textwidth]{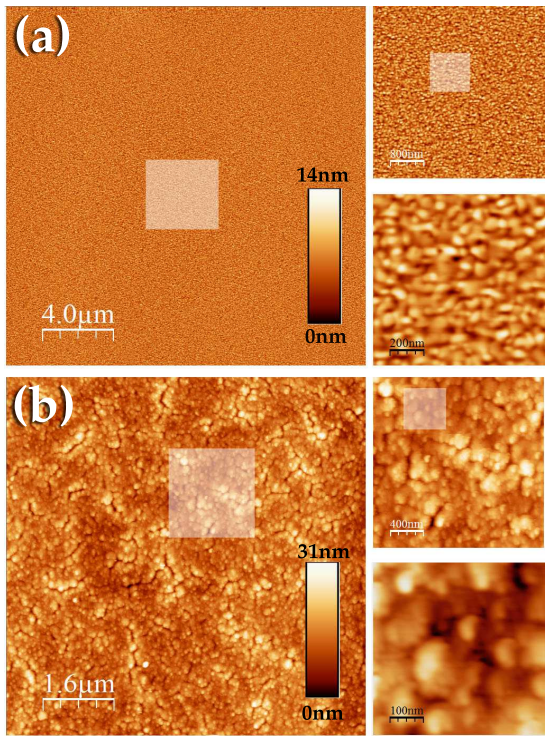}
\vspace{-0.1cm} \caption{(Color online) AFM megascan of the 100 nm
film (a) and the sphere (b). The insets show the highlighted areas
at higher magnifications. } \label{fig1} \vspace{-0.5cm}
\end{center}
\end{figure}

The surface profile was recorded with Veeco Multimode atomic force
microscope (AFM) using Nanoscope V controller. To analyze the effect
of scale dependence, megascans of large area up to $40\times 40\;\mu
m^2$ were made and recorded with the lateral resolution of
$4096\times 4096$ pixels. The maximal area, which we have been able
to scan on the sphere, was $8\times 8\;\mu m^2$ ($2048\times 2048$
pixels). All images were flattened with linear filtering; for the
sphere the parabolic filtering was used to exclude the effect of
curvature. Figure \ref{fig1} shows the images of the $100\;nm$ film
(a) and the sphere (b) on different scales. Approximately 10 images
of smaller size $500\times 500\; nm^2$ were recorded for each film
and for the sphere to obtain the correlation length $\xi$ of the
rough surfaces \cite{Pal93}. Finally, the electrostatic calibration
was used for the determination of the cantilever spring constant and
$d_0$ \cite{Zwo08a}.

\section{Plate-plate contact}
\label{sec:3}

Consider first two parallel plates, which can come into contact. A plate
surface can be described by a roughness profile $h_i(x,y)$ ($i=1,2$
for body 1 or 2), where $x$ and $y$ are the lateral coordinates. The
averaged value over large area of the profile is zero, $\left\langle
h_i(x,y)\right\rangle=0$. Then the local distance between the plates
is
\begin{equation}\label{loc_dis}
    d(x,y)=d-h_1(x,y)-h_2(x,y),
\end{equation}
where $d$ is the distance between the average planes. We can define
the distance upon contact $d_0$ as the largest distance $d=d_0$, for
which $d(x,y)$ becomes zero.

It is well known from contact mechanics \cite{Gre66} that the
contact of two elastic rough plates is equivalent to the contact
of a rough hard plate and an elastic flat plate with an effective
Young's modulus $E$ and a Poisson ratio $\nu$. In this paper we
analyze the contact in the limit of zero load when both bodies can
be considered as hard. This limit is realized when only weak
adhesion is possible, for which the dispersive forces are
responsible. Strong adhesion due to chemical bonding or due to
capillary forces is not considered here. This is not a principal
restriction, but the case of strong adhesion has to be analyzed
separately.  Equation (\ref{loc_dis}) shows that the profile of
the effective rough body is given by
\begin{equation}\label{comb}
    h(x,y)=h_1(x,y)+h_2(x,y).
\end{equation}
The latter means that $h(x,y)$ is given by the combined image of the
surfaces facing each other.

Let $L_0$ be the size of the combined image. Then, in order to
obtain information on the scale $L=L_0/2^n$, we divide this image on
$2^n$ subimages. For each subimage we find the highest point of the
profile (local $d_0$), and average all these values. This procedure
gives us $d_0(L)$ and the
corresponding statistical error. Megascans are very convenient for
this purpose otherwise one has to collect many scans in different
locations.

\begin{figure}[ptb]
\begin{center}
\includegraphics[width=0.45\textwidth]{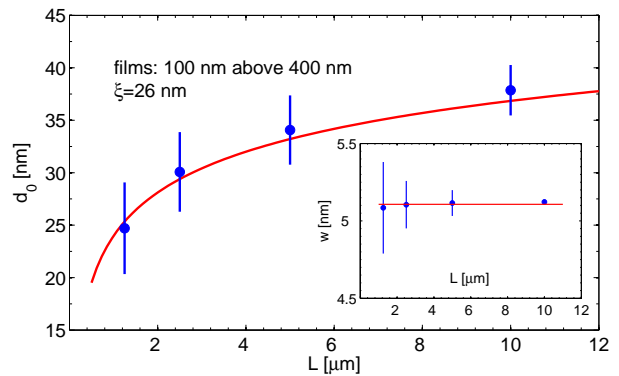}
\vspace{-0.1cm} \caption{(Color online) Distance upon contact as a
function of the length scale. Dots with the error bars are the
values calculated from the megascans. The solid curve is the
theoretical expectation according to Eq. (\ref{above_d0}). The inset
demonstrates absence of the scale dependence for the rms roughness.
} \label{fig3} \vspace{-0.5cm}
\end{center}
\end{figure}

For the $100\; nm$ film above the $400\; nm$ film the result of this
procedure is shown in Fig. \ref{fig3}. We took the maximum area to
be $10\times 10\;\mu m^2$. The figure clearly demonstrates the
dependence of $d_0$ on the scale $L$ although the errors appear to
be significant. The inset shows the dependence of the rms roughness
$w$ on the length scale $L$. This dependence is absent in accordance
with the expectations, while only the error bars increase when $L$
is decreasing.

To understand the dependence $d_0(L)$ let us assume that the size
$L$ of the area of nominal contact is large in comparison with the
correlation length, $L\gg\xi$. It means that this area can be
divided into a large number $N^2=L^2/\xi^2$ of cells. The height of
each cell (asperity) can be considered as a random variable $h$
\cite{Per06}. The probability to find $h$ smaller than some value
$z$ can be presented in a general form
\begin{equation}\label{prob}
    P(z)=1-e^{-\phi(z)},
\end{equation}
where the "phase" $\phi(z)$ is a nonnegative and nondecreasing
function of $z$. Note that (\ref{prob}) is just a convenient way to
represent the data: instead of cumulative distributions $P(z)$ we
are using the phase $\phi(z)$.

For a given asperity the probability to find its height above $d_0$
is $1-P(d_0)$, then within the area of nominal contact one asperity
will be higher than $d_0$ if
\begin{equation}\label{above_d0}
    e^{-\phi(d_0)}\left({L^2}/{\xi^2}\right)=1\ \ \ \textrm{or}\ \ \
    \phi(d_0)=\ln\left({L^2}/{\xi^2}\right).
\end{equation}
This condition can be considered as an equation for the asperity
height because due to a sharp exponential behavior the height is
approximately equal to $d_0$. To solve (\ref{above_d0}) we have to
know the function $\phi(z)$, which can be found from the roughness
profile.

\begin{figure}[ptb]
\begin{center}
\includegraphics[width=0.45\textwidth]{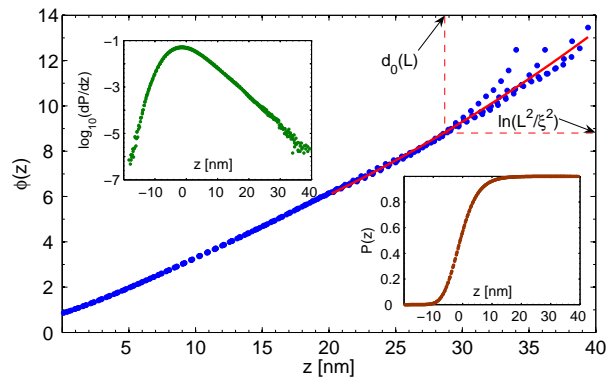}
\vspace{-0.1cm} \caption{(Color online) Statistics of the surface
roughness. Four $10\times 10\;\mu m^2$ images were used. The main
graph shows the "phase" as a function of $z$. The red (light gray)
curve is the best fit of the data at large $z$ and the dashed lines
demonstrate the solution of Eq. (\ref{above_d0}). The top inset
presents the logarithm of the density function. The bottom inset
shows the cumulative distribution. } \label{fig4} \vspace{-0.5cm}
\end{center}
\end{figure}

The cumulative distribution $P(z)$ can be found from a roughness
profile by counting pixels with the height below $z$. Then the
"phase" can be calculated as $\phi(z)=-\ln(1-P)$. The results are
presented in Fig. \ref{fig4}. It has to be noted that the function
$\phi(z)$ becomes more dispersive at large $z$. This effect was
observed for all surfaces we investigated. To solve Eq.
(\ref{above_d0}) we have to approximate the large $z$ tail of
$\phi(z)$ by a smooth curve. Any way of the data smoothing is
equally good, and our method is not relied on specific assumptions
about the probability distribution. The procedure of solving Eq.
(\ref{above_d0}) is shown schematically in Fig. \ref{fig4}, and
the solution itself is the red curve in Fig. \ref{fig3}.

It has to be mentioned that the normal distribution fails to
describe the data at large $z$. Other known distributions are not
able satisfactory describe the data at all $z$. Asymptotically at
large $z$ the data can be reasonably well fit with the generalized
extreme value distributions Gumbel or Weibull \cite{Col01,Jho94}.
This fact becomes important if one has to know $d_0$ for the size
$L$, which is larger than the maximal scan size. In this case one
has to extrapolate $\phi(z)$ to large  $z$ according to the chosen
distribution. In this paper we are not doing extrapolation using
only $\phi(z)$ extracted directly form the megascans.

The observed dependence $d_0(L)$ can be understood intuitively. The
probability to have one high asperity is exponentially small but the
number of asperities increases with the area of nominal contact.
Therefore, the larger the contact area, the higher probability to
find a high feature within this area.

Our result found in the limit of zero load will hold true if the
elastic deformation of the highest asperity will be small ($\ll
d_0$). Applying Hertzian theory to an asperity of radius $\xi/2$ one
finds the restriction on the load $p$:
\begin{equation}\label{load}
    p\ll\sqrt{2\xi/9d_0}\left(1-\nu^2\right)^{-1}\left(d_0^2/L^2\right)E.
\end{equation}
If $p=A_{H}/6\pi d_0^3$ is the van der Waals pressure ($A_H$ is
the Hamaker constant)  then (\ref{load}) for the $Au$ parameters
restricts $d_0$ and $L$ as $(d_0/10\;nm)^{4.5}(L/10\;\mu
m)^{-2}\gg 0.3$. This condition is true in the range of main
interest. For the sphere-plane case (see below) Eq. (\ref{load})
can be modified accordingly but in general the physical contact is
not assumed for the sphere-plate configuration.

\section{Sphere-plate contact}
\label{sec:4}

The other question of great practical importance is the distance
upon contact between a sphere and a plate. In the experiments
\cite{Har00,Dec03,Jou09,Man09,Zwo08a} the sphere attached to a
cantilever or an optical fibre approaches the plate. Assuming that
the sphere is large, $R\gg d$, the local distance is
\begin{equation}\label{loc_sph_2}
    d(x,y)=d+\left(x^2+y^2\right)/2R-h(x,y),
\end{equation}
where $h(x,y)$ is the combined profile of the sphere and the plate.

Again, $d_0$ is the maximal $d$, for which the local distance
becomes zero. This definition gives
\begin{equation}\label{d0_sph}
    d_0=\max\limits_{x,y}\left[h(x,y)-\left(x^2+y^2\right)/2R\right].
\end{equation}
In contrast with the plate-plate configuration now $d_0$ is a
function of the sphere radius $R$, but, of course, one can define
the length scale $L_R$ corresponding to this radius $R$ (see below).

As input data in Eq. (\ref{d0_sph}) we used the combined images of
the sphere and different plates. The origin ($x=0,\;y=0$) was chosen
randomly in different positions and then $d_0$ was calculated
according to (\ref{d0_sph}). We averaged $d_0$ found in 80 different
locations to get the values of $d_0^{im}$, which are collected in
Tab. \ref{tab1}.

\begin{table}
\begin{tabular}{l||l|l|l|l|l}
 & $100\;nm$ & $200\;nm$ & $400\;nm$ & $800\;nm$ & $1600\;nm$ \\
\hline\hline $w$ & 3.8 & 4.2 & 6.0 & 7.5 & 10.1 \\
   $\xi$ & $26.1\pm 3.8$ & $28.8\pm 3.7$ & $34.4\pm 4.7$ & $30.6\pm 2.4$ & $42.0\pm 5.5$ \\
   $L_R$ & 920 & 1050 & 1470 & 1560 & 2100 \\
   $d_0^{th}$ & 12.5 & 14.0 & 22.8 & 31.5 & 53.0 \\
   $d_0^{im}$ & $12.8\pm 2.2$ & $15.9\pm 2.7$ & $24.5\pm 4.8$
   & $31.3\pm 5.4$ & $55.7\pm 9.3$ \\
   $d_0^{el}$ & $17.7\pm 1.1$ & $20.2\pm 1.2$ & $23.0\pm 0.9$
   & $34.5\pm 1.7$ & $50.8\pm 1.3$

\end{tabular}
\caption{The parameters characterizing the sphere-film systems (all
in $nm$). The first five rows were determined from combined images
(see text). The last row $d_0^{el}$ gives the values of $d_0$
determined electrostatically. The last four rows were determined for
$R=50\;\mu m$.}\label{tab1}
\end{table}

We can estimate the same value theoretically. A circle of a finite
area $L^2$ is important in Eq. (\ref{d0_sph}). Asperities of the
size $\xi$ are distributed homogeneously within this circle. Then
the averaged value of the second term in (\ref{d0_sph}) is
$L^2/4\pi R$. The averaged maximal value of $h(x,y)$ is the
distance upon contact between two plates of the size $L$. This
distance is the solution of Eq. (\ref{above_d0}). In this section
we will denote it as $d_0^{pp}(L)$ not to mix with $d_0$ in the
sphere-plate configuration. Then one can find $d_0$ for the
sphere-plate contact by maximizing (\ref{d0_sph}) on $L$:
\begin{equation}\label{max_d0}
    d_0=\max\limits_{L}\left[d_0^{pp}(L)-L^2/4\pi R\right].
\end{equation}
The solution of this equation defines $d_0^{th}$ and the scale $L_R$
corresponding to the maximum. The values of $d_0^{th}$ and $L_R$
found from (\ref{max_d0}) are given in Tab. \ref{tab1} for the
radius $R=50\; \mu m$.

One can see that $d_0^{th}$ is in agreement with $d_0^{im}$
determined from the combined images. Comparing it with the values
$d_0^{el}$ determined electrostatically one sees that in the first
two columns the values of $d_0^{el}$ are considerably larger.
Moreover, the errors in $d_0^{el}$ are smaller than in $d_0^{im}$.

We described $d_0$ as the value determined from the area $L_R^2$ and
averaged over its different locations. Determination of $d_0$ from
the electrostatic measurements did not undergo this type of
averaging. As a result it is sensitive to the local roughness
realization near the contact location. This explains why the errors
in $d_0^{el}$ are smaller: statistical variation of $d_0$ from place
to place is not included in the errors of $d_0^{el}$.

Very different local values of $d_0$ can be found and for this
reason $d_0^{el}$ can deviate significantly from the mean value.
Choosing arbitrarily the contact locations in the image of the
sphere and the 100 nm film we found, for example, that about 5\% of
the cases are in agreement with the measured value $d_0^{el}=17.7\pm
1.1\; nm$. One can imagine that the place of contact on the sphere
has at least one asperity above the average. In the combined image
the sphere dominates since it is rougher than the film,
$w_{sph}=3.5\;nm$ and $w_{100}=1.5\;nm$. Because the sphere is
rigidly fixed on the cantilever the same feature will be in the area
of contact for any other location or other film. Already for the
sphere above 400 nm film the high feature on the sphere will not
play significant role because the roughness of the film,
$w_{400}=4.9\; nm$, is higher than that for the sphere. In this case
we would expect that $d_0^{el}$ has to be in agreement with the
averaged value found from the image that is precisely what happens.

\begin{figure}[ptb]
\begin{center}
\includegraphics[width=0.45\textwidth]{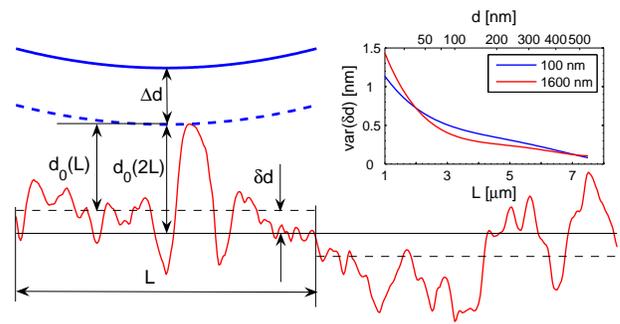}
\vspace{-0.1cm} \caption{(Color online) Schematic explanation of
additional uncertainty $\delta d$ in $d_0$ (see text). The sphere in
two positions is shown by the dashed (contact) and solid blue (dark
gray) curves. The inset shows the variance of $\delta d$ as a
function of the scale $L$ or separation $d$. } \label{fig5}
\vspace{-0.5cm}
\end{center}
\end{figure}

Consider now the experimental situation when the dispersive force is
measured in the sphere-plate configuration. The system under
consideration is equivalent to a smooth sphere above a combined
rough profile $h(x,y)$. The position of the average plane depends on
the area of averaging $L^2$ especially for small scales $L$. The
profile shown in Fig. \ref{fig5} demonstrates different mean values
in the left and right segments shown by the dashed black lines. Both
of these values deviate from the middle line for the scale $2L$
(solid black line). The true average plane is defined for
$L\rightarrow\infty$.

From Fig. \ref{fig5} one can see that $d_0$ for $L$ and $2L$ differ
on $\delta d=d_0(L)-d_0(2L)$. To be more precise we can define the
uncertainty in $d_0$ as $\delta d(L)=d_0(L)-d_0$, where we
understand $d_0$ as the value counted from the true average plane
($L\rightarrow\infty$). The distance between bodies is then
$d=d_0+\delta d(L)+\Delta d$, where $\Delta d$ is the displacement
from the contact point. The scale $L$ is defined by the effective
area of interaction $L^2=\alpha\pi Rd$ ($\alpha=2$ for the
electrostatic and $\alpha=2/3$ for the pure Casimir force). Suppose
that $d_0$ found from the electrostatic calibration can be
considered as a true value (the electrostatic scale is large,
$L_{el}\rightarrow\infty$) then in the dispersive force measurement
the bodies are separated by $d=d_0+\delta d(L_{dis})+\Delta d$ with
the related scale $L_{dis}=\sqrt{\alpha\pi Rd}$.

For a fixed $L$ the uncertainty $\delta d$ is a random variable
distributed roughly normally around $\delta d=0$. However, it has to
be stressed that $\delta d$ manifests itself not as a statistical
error but rather as a kind of a systematic error. This is because at
a given lateral position of the sphere this uncertainty takes a
fixed value. The variance of $\delta d$ is defined by the roughness
statistics. It was calculated from the images and shown as inset in
Fig. \ref{fig5}. One has to remember that with a probability of 30\%
the value of $\delta d$ can be larger than that shown in Fig.
\ref{fig5}.

\section{Conclusions}
\label{sec:5}

In conclusion, it is shown that the distance upon contact depends on
the lateral size of contacting plates and a simple formula
describing $d_0(L)$ is proposed. For the sphere and plate an
additional uncertainty in the absolute separation $d$ is revealed
arising due to variation of the average plane position with the
effective area of interaction or equivalently with the separation.
Its magnitude depends on the roughness of interacting bodies.

\acknowledgments{We acknowledge helpful discussions with S.
Lamoreaux and R. Onofrio. The research was carried out under
project number MC3.05242 in the framework of the Strategic
Research programme of the Materials innovation institute M2i (the
former Netherlands Institute for Metals Research (NIMR)). The
authors benefited from exchange of ideas by the ESF Research
Network CASIMIR .}


\end{document}